\documentstyle[titlepage,12pt]{article}

\title{\bf Bit Level Correlations in Some Pseudorandom Number Generators}

\author{
K. Kankaala$^{a,b}$, T. Ala-Nissila$^{a,c}$, and I. Vattulainen$^{a,c}$
\thanks{The order of the authors' names was determined using R250.}
\\ \\
$^a$Department of Electrical Engineering \\
Tampere University of Technology \\
P.O. Box 692, FIN - 33101 Tampere \\
Finland
\\ \\
$^b$Centre for Scientific Computing \\
P.O. Box 405, FIN - 02101 Espoo \\
Finland
\\ \\
$^c$Research Institute for Theoretical Physics \\
P.O. Box 9 (Siltavuorenpenger 20 C) \\
FIN - 00014 University of Helsinki \\
Finland
\\ \\
}

\date{August 24, 1993}

\begin{document}

\maketitle

%%%%%%%%%%%%%%%%%%%%%%%%%
% 	ABSTRACT        %
%%%%%%%%%%%%%%%%%%%%%%%%%
\begin{abstract}

We present results of extensive bit level tests on some pseudorandom
number generators which are commonly used in physics applications.
The generators have first been tested with an extended version of
the $d$-tuple test. Second, we have developed a novel {\it cluster
test} where a physical analogy of the binary numbers with the two dimensional
Ising model has been utilized.
We demonstrate that the new test is rather powerful
in finding periodic correlations on bit level.
Results of both test methods are presented for each bit of the
output of the generators. Some generators exhibit clear bit level correlations
but we find no
evidence of discernible correlations for generators, which have
recently produced systematic errors in Monte Carlo simulations.

\bigskip

PACS numbers: 02.70.Lq, 05.50.+q, 75.40Mg.
\end{abstract}

\textheight 21cm
\textwidth 14.5cm
\oddsidemargin 0.96cm
\evensidemargin 0.96cm
\topmargin -0.31cm
\raggedbottom
\def\step{\: \: \:}
\baselineskip 24pt
\parindent=10mm
%\input epsf
%\epsfverbosetrue

%%%%%%%%%%%%%%%%%%%%%%%%%%%%%
%       INTRODUCTION        %
%%%%%%%%%%%%%%%%%%%%%%%%%%%%%

Vast amounts of random numbers are needed in several applications such as
stochastic optimization \cite{Aar89} and Monte Carlo simulations \cite{Bin92}.
Modern high speed computers have set rigorous demands for
the quality of random numbers, which are usually produced by pseudorandom
number generator algorithms. A prerequisite to the success of
the methods is the quality of randomness of the
output of the generators. It
is usually determined by statistical tests \cite{Knu81}.
Usually many such tests are needed, since
there is no unique recipe for determining when a given sequence is ``random
enough''.

\par
Unfortunately, even comprehensive statistical testing cannot guarantee that
a given random number generator is reliable for all applications.
In fact, tests are needed which would be more {\it physical}, based on
the use of generators in solving actual physical problems. A few such
application specific
tests have been performed \cite{Kir81,Kal84,Pau84,Mil86}.
In particular, intriguing results
have been reported by
Ferrenberg {\it et al.} \cite{Fer92} who employed some of the most commonly
used random number generators in simulations of the two dimensional Ising
model at criticality. When using the Wolff algorithm \cite{Wol89}
they reported anomalously large errors with a particular generator,
called R250. The same conclusion has been drawn from simulations of
self-avoiding random walks \cite{Gra93}, where also other similar
generators failed.

\par
Although there have been prior warnings against the use of shift
register generators such as R250 \cite{Mar85a,Mar85b},
the results of Refs. \cite{Fer92,Gra93}
are surprising since recent extensive statistical tests have found
no discernible correlations in R250 \cite{Vat93}.
In Ref. \cite{Fer92}, the authors suggest that bit level correlations in
the most significant bits of R250 may be responsible for their results.
If true, this casts serious doubt on the bit level reliability of R250.
More and also better tests are then needed to resolve the issue.
\par

The purpose of the present work is to study bit level correlations
in some commonly used generators in more detail.
To this end we have first extended the $d$-tuple test
\cite{Mar85b,Alt88} to more efficiently find correlations.
Second, we have developed a new physical {\it cluster test}
which is based on an analogy to the Ising model.
The test is implemented on
bit level and its effectiveness compared with the $d$-tuple test.
We demonstrate that the cluster test is particularly powerful in finding
periodic correlations. Both the $d$-tuple and the cluster tests
are then applied to each bit of a number of generators, including
some of the shift register generators in Refs. \cite{Fer92,Gra93}.
Our results demonstrate that no discernible bit level correlations
can be found in the shift register generators
with the present test methods.
%
%%%%%%%%%%%%%%%%%%%%%%%%%%%%%
%	INTRO OF RNGS       %
%%%%%%%%%%%%%%%%%%%%%%%%%%%%%
\par
The pseudorandom number generators used here include two linear
congruential generators, LCG(16807,0,$2^{31}-1$) \cite{Par88}
known as GGL (CONG in Ref. \cite{Fer92}),
and LCG(69069,1,$2^{32}$) \cite{Mar72} implemented as RAND
\cite{convex}. Additionally, RAN3 \cite{Pre89} is a LF(55,24,$-$) based on a
Lagged Fibonacci algorithm, whereas RANMAR \cite{Jam90,Mar90a}
is a combination generator. Finally, GFSR(250,103,$\oplus$) and
GFSR(1279,216,$\oplus$) are generalized feedback shift register
generators known as R250 \cite{Kir81,Lew73} and R1279,
respectively. The details of the algorithms can be found,
for example, in Ref. \cite{Vat93}. We note that the generators were implemented
to produce integers except for RANMAR, whose 24 bit reals were multiplied
by $2^{31}-1$. Initial seed values were chosen
from the set $\{14159, 667790, 1415926535, 95141\}$,
excluding R250 and R1279 which were initialized
with GGL in double precision accuracy.

%%%%%%%%%%%%%%%%%%%%%%%%%%%%%%%%
%	EXTENDED D-TUPLE       %
%%%%%%%%%%%%%%%%%%%%%%%%%%%%%%%%
\par
The $d$-tuple test is based on studying the properties of random
numbers on bit level \cite{Mar85b}. Our realization follows Ref.
\cite{Alt88}. The main
difference here is the improvement
to calculate the $\chi^2$-distributed test statistics a total of
$N$ times and submit their empirical distribution to a Kolmogorov -
Smirnov (KS) test. The final test variables are therefore the values
$K^+$ and $K^-$ of a KS - test statistic $K$ \cite{Knu81}. In each
test the sequence of bits was considered to fail if the
observed descriptive level ${\delta}
 = P(K \leq \{ K^+, K^-\} | H_0)$ was less than 0.05 or larger than 0.95.

\par
Based on previous work \cite{Vat93} the $d$-tuple test seems to find
correlations more efficiently than the rank test \cite{Mar85a,Mar85b},
for example. In order to determine the quantitative effectiveness of
the test we have first studied its ability to observe correlations
inserted into the output of GGL, which passes the standard
bit tests \cite{Vat93}.
The correlations have been inserted
periodically by setting the $i^{\rm th}$ bit ($i=1,2,\ldots,31$) of every
$\xi^{\rm th}$ number always equal to one.
By systematically varying $\xi$, we can then find the maximum approximate
distance $\xi_c$
within which the $d$-tuple test can detect periodic correlations.
The test was repeated
three times with parameters $d=t=3$, $n=5000$ and $N=1000$, where $d$
and $t$ are taken from Refs. \cite{Vat93,Alt88}
and $n$ is the number of samples
in a single $\chi^2$ - test. The results are shown in Table 1, where
the parameter $p$ gives the probability of observing correlations.
Thus, the $d$-tuple test can always detect periodic correlations up to
$\xi_c \approx 43$ bits apart. The same test was repeated with $d=9$
and $t=1$ to consider single bits only, which gave $\xi_c \approx 50$.
We also note that we performed similar systematic tests for the rank
test, which was found to be inferior to the $d$-tuple test.
\par

To improve the detection range of the $d$-tuple test we have performed
its {\em extended version}. This can be realized by testing bits
from every $k^{\rm th}$ number and then testing all $k$ such
subsequences. This way all periodic
correlations may be detected up to about $k \times \xi_c$ (assuming
$k < \xi_c$).
We have applied this extended test to GGL, R250, R1279, and RAN3,
which were all tested twice. The results are summarized in Table 2.
The most remarkable result is that up to $k=20$, which corresponds to
a distance of about $860$ bits apart, no discernible correlations
were observed for the 16 most significant bits of either R250 or R1279.
In addition, we tested R250
with $k=50$, $k=100$ and $k=1000$ where only one
subsequence was studied in each case. No evidence of correlations was found.
This result is in contrast with Ref. \cite{Gra93}, where it was
estimated that for R250 a typical range of correlations is about 400.
However, when initialized with
RAN3 which itself contains correlated bits, both R250 and R1279 display
clear bit level correlations, although the longer feedback of R1279
seems to be less sensitive to initial correlations.

%%%%%%%%%%%%%%%%%%%%%%%%%%%%%%%%%%%
%	THE CLUSTER TEST          %
%%%%%%%%%%%%%%%%%%%%%%%%%%%%%%%%%%%
\par
There is a natural analogy between binary numbers and the
Ising model, which can be made use of in constructing
a new physical {\it cluster test} in the following way.
We take $i^{\rm th}$ bit from every successive number and put them
on a two dimensional lattice of size $L^2$.
By identifying zeros and ones with
the ``down'' and ``up'' spins of the Ising model, the resulting configuration
--- if truly random --- should be one of the $2^{L^2}$ equally weighted
configurations corresponding to infinite temperature.
The easiest quantity that one can then compute from this analogy is the
magnetization.
However, a better measure of {\it spatial} correlations
can be obtained if we study the distribution of connected spins, or
clusters of size $s$ on the lattice.
The cluster size distribution
$\langle n_s \rangle$ is given by \cite{Syk76}
\begin{equation}
\langle n_s \rangle = s p^{s} D_{s}(p),
\label{Eq:n_s}
\end{equation}
where $D_s (p)$'s are polynomials in $p=1/2$.
The normalization condition is
$\sum_{s=1}^{\infty} \langle n_s \rangle = 1$.
Enumeration of the polynomials $D_s (p)$ has been done up to
$s=17$ \cite{Syk76}.

\par
The test procedure we have used is as follows.
We first form a $L^2$ lattice as above and enumerate
all the clusters in it.
For such a configuration we calculate the (unnormalized) average size
of clusters within $s = 1, 2, \ldots, 17$, denoted as $S_{17}^{(k)}$.
This procedure is repeated
$M$ times corresponding to configurational averaging,
yielding $S_{17}= \sum_{k=1}^M S_{17}^{(k)}/M$.
The theoretical counterpart of this quantity is given by
$ s_{17} = \sum_{s=1}^{17} s \langle n_s \rangle$.
We also compute the empirical standard deviation
$\sigma_{17}$ of the quantities $S_{17}^{(k)}$.
For each $i^{\rm th}$ bit the test statistic chosen in this work is:
\begin{equation}
g_i =
\frac{S_{17} - s_{17}}{\sigma_{17}}.
\end{equation}
Using this statistic, tests were performed
comparatively between several pseudorandom number generators, with
results from GGL assumed to be independent variables \cite{Gaussian}.
Therefore, the mean value of $g_i$ over all the 31 bits of GGL, denoted
as $g_{\rm GGL}$ and the corresponding standard
deviation $\sigma_{\rm GGL}$ were computed and the results
for all other generators were compared with these
values using
\begin{equation}
g_i'=
\frac{\vert g_i - g_{\rm GGL} \vert }{\sigma_{\rm GGL}}.
\end{equation}
The bit $i$ in question
failed the test if $g_i'$ was greater than
three. We also considered other similar choices for the test
parameters and criteria and
obtained consistent results.
\par

The effectiveness of the cluster test was first scrutinized
by inserting periodic correlations as in the case of the $d$-tuple test.
We chose $L=200$, $M = 10000$ and the study was repeated for
all values of $\xi = 1, 2, \ldots, L$. The results are shown in Table 3,
where filled squares denote distances where correlations were detected.
With this choice of parameters the cluster test is able to find all
periodic correlations up to $\xi_c \approx 60$.
This shows that the cluster test performs somewhat better than either
the $d$-tuple or rank tests.
\par

Next, we have subjected each bit of
the random number generators to the cluster test.
It was repeated twice with parameters $L=200$, and $M=10000$. Additional
tests with $L=500$ gave consistent results.
Results are summarized in Table 4, where also
results of the previous $d$-tuple and rank tests from Ref. \cite{Vat93}
have been included.
Although more powerful than the other methods, the cluster test still
reveals no discernible correlations for either GGL, R250 or R1279.
For RANMAR and RAN3, the cluster test gives results consistent with
Ref. \cite{Vat93}, but for RAND additional correlations
are revealed in bits 8 - 12, which passed the $d$-tuple test.

\par
For completeness, we also tested the distribution of bits.
The bits failed the test if the deviation
from the expected number of ones (i.e. $L^2$/2)
consequtively exceeded three times
the standard deviation of the binomial distribution with $M$ samples.
The test was repeated twice with $M = 4 \times 10^8$ and its
results are also shown in Table 4.
No correlations were found for GGL, R250, or R1279.
Surprisingly, however,
this rather simple test revealed that the first 11 bits of
RAN3 fail (with standard deviations larger than 6.7)
although only the first four
or five bits fail in the other tests. This signals
correlations in these additional bits.
On the other hand, for RAND only bits 22 - 31 failed,
which produced an exact 50 - 50 distribution of zeros and ones.

%%%%%%%%%%%%%%%%%%%%%%%
%       SUMMARY       %
%%%%%%%%%%%%%%%%%%%%%%%
\par
In conclusion, we have performed extensive bit level tests
of several commonly used pseudorandom number generators, including
R250 which had been suggested to contain bit level correlations
\cite{Fer92}. To this end, we have performed an extended version
of the $d$-tuple test, and developed a novel physical {\it cluster test},
which is rather powerful in finding periodic correlations.
Our results reveal significant bit level correlations in some
generators, such as RAN3 and RAND, but absolutely no discernible
correlations in GGL, R250, or R1279. Thus, our results
show that these generators should be good enough for many applications,
where good bit level properties are required.
However, we note that it is still
of crucial importance to further develop physical
tests along the lines presented here to detect more subtle correlations,
which may not be revealed by the present test methods.

\pagebreak

\Large
{\bf Table Captions}
\normalsize
\baselineskip 24pt
\parindent=0mm
\bigskip

Table 1. Results of the $d$-tuple test with inserted
         correlations in the bits, with a
         period of $\xi$. The probability for the test to observe
         correlations is denoted by $p$,
         which equals one up to $\xi_c \approx 43$.

\bigskip

Table 2. Results of the extended $d$-tuple tests. $k$ denotes the extended
         range of the tests. See text for details.

\bigskip

Table 3. Results of the cluster test with correlations in the bits, with a
         period of $\xi$ from one to 200.
         Black squares denote corresponding
         distances at which correlations were
         found as explained in the text.

\bigskip

Table 4. Results of the cluster test. $d$-tuple and rank test results are
         from Ref. \cite{Vat93}. The last column denotes bits which fail
         in testing the distribution of ones.

\pagebreak
%%%%%%%%%%%%%%%%%%%%%%%%%
%      BIBLIOGRAPHY     %
%%%%%%%%%%%%%%%%%%%%%%%%%
Electronic addresses:
{\tt Kari.Kankaala@csc.fi, ala@phcu.helsinki.fi, Ilpo.Vattulainen@csc.fi}
\par

\pagebreak
% ==========================================================================
%
%	Table 1: Results of GGL with artificial correlations.
%		 d-tuple and rank tests.
%
% ==========================================================================
\begin{table}[htb]
\normalsize \centering
\begin{tabular}{| c | c | c | c | c | c | c | c | c | c | c |}
%\hline
%\multicolumn{11}{| c |}{$d$-tuple test} \\
\hline
$\xi$ 	& 40 	& 43 	& 52 	& 60 	& 70
	& 80 	& 90 	& 100	& 110 	& 120	\\ \hline
$p$ 	& 1.000	& 0.889	& 0.778	& 0.333	& 0.667
	& 0.222	& 0.333	& 0.111	& 0.222 & 0.000 \\
\hline
\end{tabular}
\caption{}
\end{table}

% ==========================================================================
%
%	Table 2: Results of R250 and R1279 with two types of initializations.
%
% ==========================================================================
\begin{table}[htb]
\normalsize \centering
\begin{tabular}{| l | c | l | l |}
\hline
Random 		& 	& Failing bits		&  \\
number 		& $k$ 	& in the 		& Comments \\
generator 	& 	& {\it d}-tuple test 	&  \\
\hline\hline
GGL 		& 1,5	& none 		& double precision mode\\
        	&	& 		& (return integers)\\
\hline
R250 		& 1,5	& none 		& integer mode, initialized\\
        	&	&		& with GGL in double precision\\
R250 		& 20	& none 		& only 16 most significant \\
        	&	&		& bits were studied\\
\hline
R1279 		& 1,5	& none 		& integer mode, initialized\\
        	& 	& 		& with GGL in double precision\\
R1279 		& 20	& none 		& only 16 most significant \\
        	&	&		& bits were studied\\
\hline\hline
RAN3 		& 1	& $1-5$, $25-30$	& integer mode\\ \hline
R250 		& 1	& $1-2$, $27-31$	& integer mode, initialized\\
        	&	&		& with RAN3 producing integers\\
\hline
R1279 		& 1	& $1$		& integer mode, initialized\\
        	&	&		& with RAN3 producing integers\\
\hline
\end{tabular}
\caption{}
\end{table}

\clearpage

% ==========================================================================
%
%	Table 3: Results of GGL with artificial correlations.
%		 The cluster test.
%
% ==========================================================================

\begin{table}
%\vspace{-20.0cm}
%\hspace{-5cm}
%\epsfbox[0 0 850.43 1202.71]{cl_cr2.ps}
%\special{psfile=cl_cr2.ps angle=270 hscale=80 vscale=80
%hoffset=50 voffset=200}
%\vspace{5.0cm}
\vspace{15cm}
\caption{}
\end{table}

\clearpage

% ==========================================================================
%
%	Table 4: The summary of the results of the bit level tests.
%	         The results of the cluster test are included.
%
% ==========================================================================
\begin{table}[h]
\normalsize \centering
\begin{tabular}{| l | c | c | c | c |}
\hline
\multicolumn{1}{| c |}{Random} & \multicolumn{4}{ c |}{Failing bits} \\
\cline{2-5}
\multicolumn{1}{| c |} {number}
& \multicolumn{1}{c |}{Cluster test}
& \multicolumn{1}{c |}{$d$-tuple test}
& \multicolumn{1}{c |}{Rank test}
& \multicolumn{1}{c |}{Distribution} \\
\multicolumn{1}{| c |}{generator}
& \multicolumn{1}{c |}{} 	& \multicolumn{1}{c |}{}
& \multicolumn{1}{c |}{} 	& \multicolumn{1}{c |}{of bits}
\\ \hline\hline
GGL	& none 			& none 			& none
	& none 	\\
R250	& none 			& none 			& none
	& none 	\\
R1279	& none 			& none 			& none
	& none 	\\
RANMAR	& $25-31$ 		& $25-31$ 		& $25-31$
	& $25-31$ \\
RAN3	& $1-4$, $25-30$	& $1-5$, $25-30$	& $1-5$, $26-30$
	& $1-11$, $24-30$ \\
RAND	& $8-31$		& $13-31$		& $18-31$
	& $22-31$ \\
\hline
\end{tabular}
\caption{}
\end{table}


\begin{thebibliography}{250}
\bibitem{Gaussian}
We checked numerically that to a good degree of approximation, the
distribution for $S_{17}^{(k)}$ is given by a Gaussian for GGL.
\end{thebibliography}
\end{document}